\documentclass[prb,twocolumn,epsfig,epsf]{revtex4}
\usepackage{amsmath,amsthm,amsfonts,amscd,mathrsfs,pifont,bm}
\usepackage{eucal,graphicx,color}
\usepackage{esint}

\makeatletter
\@for\next:={int,iint,iiint,iiiint,dotsint,oint,oiint,sqint,sqiint,
  ointctrclockwise,ointclockwise,varointclockwise,varointctrclockwise,
  fint,varoiint,landupint,landdownint}\do{%
    \expandafter\edef\csname\next\endcsname{%
      \noexpand\DOTSI
      \expandafter\noexpand\csname\next op\endcsname
      \noexpand\ilimits@
    }%
  }
\makeatother

\newcommand{\beq}{\begin{equation}}
\newcommand{\eeq}{\end{equation}}

\newcommand{\M}[1]{\pmb{#1}}

\begin{document}

\title  {Universal Collective Behavior of the Vibration Spectrum of Highly Connected Disordered Systems}

\author {Joshua Feinberg\footnote{ https://orcid.org/0000-0002-2869-0010} and Roman Riser}

\affiliation
       {Department of Mathematics\\ and\\
Haifa Research Center for Theoretical Physics and Astrophysics University of Haifa, Haifa 31905, Israel}

\date   {December 10, 2020}

\begin  {abstract}
We study small oscillations of highly connected systems, which represent a limit opposite to the more familiar case of disordered crystals. As a concrete example we analyze the vibrational spectra of composite pendula. Remarkably, these spectra exhibit universal behavior with non-vanishing zero-frequency limit of the density of phonon states. This universality is captured by a natural random matrix model for such systems. We analyze this model using S-transforms of free probability theory and obtain the density of modes explicitly, in the limit of large system size.   
\end{abstract}

\maketitle
{\it Introduction.} The problem of determining the small oscillations of a mechanical system about a stable equilibrium state is ubiquitous in physics. Thus, given a system with $N$ degrees of freedom and corresponding  generalized coordinates ${\bf q} = (q_1,\ldots q_N)$, its small oscillations about a stable equilibrium point ${\bf q}_0$ are solutions of the linearized equations of motion \cite{LL, Arnold, GK}
\begin{equation}\label{small-oscillations}
\M{M} \ddot {\bf x} +  \M{K} {\bf x}  = 0\,, 
\end{equation}
where the $N\times N$ {\em strictly positive-definite} matrix $\M{M} = \M{a}({\bf q}_0)$ is the value of the metric appearing in the kinetic part the lagrangian $L = \frac{1}{2}\dot{\bf q}^T \M{a}({\bf q}) \dot{\bf q} - U({\bf q})$ evaluated at ${\bf q}_0$, the {\em positive} matrix $K_{ij} = {\partial^2 U\over \partial q_i\partial q_j}({\bf q}_0)$ is the Hessian of the potential at the equilibrium point, and ${\bf x} = {\bf q}-{\bf q}_0$ is a small deviation from equilibrium. Harmonic eigenmodes of the system are solutions of the form ${\bf x}(t) = {\bf A} e^{i\omega t}$, for which \eqref{small-oscillations} implies the characteristic equation 
\begin{equation}\label{eigen}
\left(-\omega^2 \M{M} + \M{K}\right){\bf A} = {\bf 0}
\end{equation}
for the eigenvector amplitude and frequency eigenvalue. 
Eigenfrequencies are roots of the characteristic polynomial $P_N(\omega^2)= \det \left(-\omega^2 \M{M} + \M{K}\right)$. These roots are all positive, since according to \eqref{eigen}, $\omega^2 ={{\bf A}^\dagger \M{K}\bf{A}\over {\bf A}^\dagger \M{M}{\bf A}}$ is the ratio of two positive quantities. This should be expected on physical grounds, since in the absence of dissipation, the eigenfrequencies $\omega$ of a stable system are all real.

Under certain conditions, the problem of small oscillations \eqref{small-oscillations}-\eqref{eigen} naturally lends itself to analysis by means of random matrices. One very important example is the determination of the average phonon spectrum of disordered crystals. This problem was solved long ago in one-dimension by Dyson \cite{Dyson} and Schmidt \cite{Schmidt}. Individual ions in a disordered chain are subjected to nearest-neighbor interactions. Consequently, the mass matrix $\M{M}$ is diagonal, and the spring-constant matrix $\M{K}$ is tri-diagonal (a Jacobi matrix). Randomness arises either due to having random ion masses on the diagonal of $\M{M}$, or random spring constants in $\M{K}$ (or due to both). 

Disordered mechanical systems with high connectivity represent a limit opposite to the more familiar case of disordered crystals. Such systems have all their degrees of freedom coupled to each other. Each momentum is typically coupled to all other momenta, and similarly for the coordinates. Thus, the problem of small oscillations in such systems involves two full positive-definite matrices $\M{M}$ and $\M{K}$. 
We can rewrite the eigenmode equation \eqref{eigen} as $\M{H}{\bf A} = \omega^2 {\bf A}$, where the ``hamiltonian" matrix is
\begin{equation}\label{H}
\M{H} = \M{M}^{-1}\M{K}\,\quad{\rm and}\quad \M{H}^\dagger = \M{K}\M{M}^{-1}\,.
\end{equation}
In general, $[\M{M},\M{K}]\neq 0$, and consequently $\M{H}^\dagger\neq \M{H}$. However, $\M{H}$ and its adjoint fulfil the {\em intertwining relation} $\M{H}^\dagger \M{M} = \M{M}\M{H}$, rendering $\M{H}$ a {\it quasi-hermitian} matrix \cite{talks,FR}.

Such systems may be inherently random (due to a random metric $\M{a}({\bf q})$ or potential $U({\bf q})$ in the lagrangian), or following Wigner's original introduction of random matrix theory into nuclear physics, may be just approximated by random matrices due to high structural complexity of the system under study.

{\it Composite Pendula.} As a concrete physical realization of the latter possibility, consider a multi-segmented pendulum made of $N$ rigid (massless) segments of lengths $l_1, \ldots l_N$ and point masses $m_1, \ldots m_N$.  The mass $m_k$ is attached to the frictionless hinge connecting segments $l_k$ and $l_{k+1}$, and carries electric charge $Q_k$. The charges $Q_1,\ldots Q_N$ are all like-sign, say positive, rendering all Coulomb interactions repulsive. The pendulum is suspended by the other end of the  first segment $l_1$ from a frictionless hinge, which is fixed to an infinite mass $m_0$ (a wall), carrying positive charge $Q_0$. The whole system is suspended in Earth's gravity $g$, and is free to execute planar oscillations. (This system can be thought of as a model for a charged (unscreened) polymer chain in a uniform external field.)

Let $ \theta_k\in[-\pi,\pi]$ be the angle between the segment $l_k$ and the downward vertical. Clearly, the stable equilibrium state of the system occurs when all segments align vertically, i.e.~when all $\theta_k=0$. We have studied the small oscillations of this pendulum about its equilibrium state, i.e.~motions for which all $|\theta_k|\ll 1$. The lagrangian governing these small oscillations is $L=\frac{1}{2} \dot{\theta}^T \M{M} \dot{\theta}-\frac{1}{2}\theta^T \M{K} \theta$, with $\theta=(\theta_1,\ldots, \theta_N)^T$, and
where the entries of the symmetric matrices $\M{M}$ and $\M{K}$ are given by
\begin{align*}
M_{ij}&=l_i l_j \!\!\!\!\!\! \sum_{~~~k=\max(i,j)}^{N}\!\!\!\!\!\!m_k,  &K_{ij}&=U_{ij}+\delta_{ij} l_i \,g \sum_{k=i}^{N} m_k,\nonumber\\
U_{ij}&=\left\{ \begin{array}{ll}
-\tilde{U}_{ij}, & \text{if $i\neq j$,}\\ \sum\limits_{\substack{k=1 \\ k\neq i}}^N\tilde{U}_{ik}, & \text{if $i=j$,}
\end{array} \right.\!\!\!\!\!\! \nonumber\\
\tilde{U}_{ij}&=l_i l_j \!\!\!\sum_{k=1}^{~\min(i,j)}  \!\!\!\!\!\!\!\!\!\sum_{~~l=\max(i,j)}^N  \!\!\!\!\!\! Q_{kl}, &Q_{ij}&=Q_{i-1} Q_{j}\left(\sum_{k=i}^{j}l_k \right)^{\!\!\!-3}.
\end{align*}
The matrix $\M{U}$ originates from the Coulomb potential energy. Note that the sum of entries in each row of this matrix vanishes. This is simply a manifestation of translational invariance of the Coulomb interaction: It depends only on squares of differences of angles $(\theta_i-\theta_j)^2$ (which arise from expanding $\cos(\theta_i-\theta_j)$ to the first nontrivial order). Therefore, shifting {\em all} $\theta_i$ by the same amount $\theta_0$ cannot change the Coulomb energy of the system. 
In other words, 
$\theta=(\theta_0,\ldots \theta_0)^T$ must be a null eigenvector of the matrix $\M{U}$, which is indeed the case. (The gravitational potential energy breaks this translational symmetry, of course.) Moreover, all other eigenvalues of $\M{U}$ should be positive, since any distortion of the pendulum from a straight line configuration will cost energy due to Coulomb repulsion. Indeed, $\M{U}$ is a diagonally dominant matrix, and is therefore positive semi-definite by virtue of Gershgorin's theorem\cite{Gershgorin}. 
If all charges and masses are non-zero, $\M{M}$ and $\M{K}$ are full matrices (the latter is due to the long-range Coulomb interaction), while in the pure gravitational model, for which $Q_0=\ldots=Q_N=0$, the matrix $\M{K}$ is diagonal.

\begin{figure}[tb]
\includegraphics[width=\columnwidth]{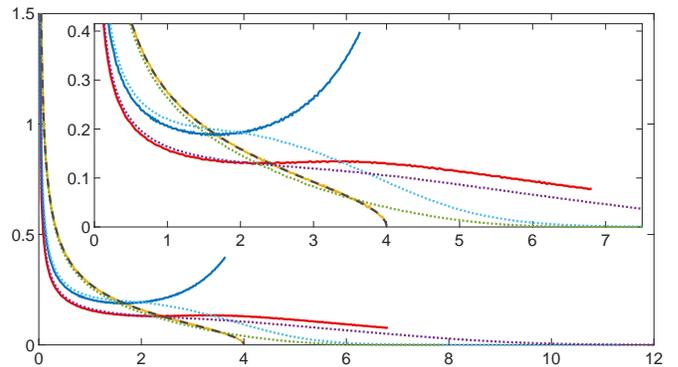}
\caption{Histogram envelope curve showing the density of states (times $N^2$) for the multi-segmented pendulum as a function of $\omega^2/N^2$. Solid lines correspond to uniform systems with all lengths, masses and charges being equal and set to $l_k=m_k=1/N$ and $Q_k={\cal Q}/(N \sqrt{\log N})$ with $N=16384$. The three graphs correspond to ${\cal Q}=1, g=1$ (red), ${\cal Q}=0, g=1$ (yellow), and ${\cal Q}=1, g=0$ (blue). The dotted lines in purple, green and cyan show the analogous curves respectively, when lengths, masses and charges are random, and with $N=1024$ averaged over $25000$ samples. The dashed black line overlying the yellow line shows the Marchenko-Pastur distribution. The inset here (and in all figures below) shows a magnification of the edge behavior for the same data.}\label{fig:pendulum}
\end{figure}

In order to get oriented, let us consider first the uniform pendulum, for which all lengths, masses and charges are equal. As we want our system to have finite total length $L$ and mass $M$ when $N\rightarrow \infty$, segment lengths and point masses must scale like $l_k\sim N^{-1}$ and $m_k\sim N^{-1}$. Under such circumstances the diagonal elements $U_{ii}$ (for segments $1 \ll i \ll N $ in the bulk of the pendulum) grow like $\log N = \log (L/l)$, because it is essentially the electrostatic potential at a point somewhere on a uniformly charged rod  of length $L$\cite{Sommerfeld}. Therefore, we have to scale the charges like $Q_k\sim (N \sqrt{\log N})^{-1}$, in order to balance the Coulomb potential energy against gravitational potential energy. Under this scaling of parameters, careful inspection of the matrices $\M{M}$ and $\M{K}$ shows that $\M{H} = \M{M}^{-1}\M{K}$ scales like $N^2$, rendering its eigenvalue bandwidth scaling likewise. We have investigated such a system numerically. The solid lines in Fig.~\ref{fig:pendulum} represent histogram envelopes for $\varrho_{\M{H}}(\omega^2)$, the (normalized) density of states (eigenvalues of $\M{H}$), as a function of $\omega^2/N^2$. These lines correspond to the cases of a mixed system with both Coulomb and gravitational interaction, a system with purely Coulomb interaction, and a system with purely gravitational interaction.  
Remarkably, as $\omega\rightarrow 0$, the density $\varrho_{\M{H}}(\omega^2)$ in all three cases diverges universally as 
\begin{equation}\label{universal}
\varrho_{\M{H}}(\omega^2) \sim {c\over\sqrt{\omega^2}} = {c\over\omega}
\end{equation} 
(with coefficients $c\neq 0$ which vary from case to case). This means that the density of frequency eigenmodes $\tilde\rho(\omega)  = 2\omega \varrho_{\M{H}}(\omega^2)$ tends to a constant $\tilde\rho(0) = 2c$ in this limit. (In the cases where there are Coulomb interactions, our numerical investigation showed slow convergence at the hard ($\omega=0$) edge which we attribute to finite-$N$ effects.)

The spectrum of the mixed system is noticeably broader due to the combined Coulomb and gravitational forces acting on the masses. While the density of states in the purely gravitational model  vanishes at the right edge of the spectrum like a square root, it exhibits a band-end discontinuity when Coulomb interactions are involved. Such discontinuous behavior of the density of states occurs also in other and completely different physical systems, such as Bloch electrons in a perfect crystal. 

Finally, an utterly surprising observation, demonstrated by the coincidence of the yellow and black dashed lines in Fig.~\ref{fig:pendulum}, is that the density of eigenvalues of the purely gravitational and completely deterministic system follows the Marchenko-Pastur distribution \cite{Marchenko-Pastur}.

We have also studied the disordered pendulum, with lengths $l_k$, masses $m_k$ and charges $Q_k$ all being i.i.d. random variables, drawn from probability distributions chosen such that the mean values of these variables would coincide with the corresponding values of these parameters in the uniform, undisordered systems displayed in Fig.~\ref{fig:pendulum}, and with standard deviations $\sigma$ of the same order of magnitude at large-$N$ as those averages. 
The resulting averaged densities of states are displayed by the dotted lines in Fig.~\ref{fig:pendulum}. 
Note that these densities of the disordered systems vanish at the high-frequency edge of the spectrum for all three cases. The band-end spectral discontinuities of some of the ordered uniform systems are smoothed out as a result of averaging over disorder. 

More importantly, note that disorder does not change the universal low-frequency divergence \eqref{universal}. Evidently, low-frequency oscillations correspond to long-wavelength collective motions, which probe the large-scale structure of the pendulum, averaged over many random segments.  Spectra of the disordered systems start to deviate from their uniform system counterparts only as the frequency increases, and oscillations become sensitive to the smaller scale structure of the system.

{\it Random Matrix Model.} The matrices $\M{M}$ and $\M{K}$ corresponding to small oscillations of the pendula discussed previously are full and rather complicated functions of the $3N$ parameters entering the problem. Such highly connected systems clearly lend themselves to analysis in terms of random matrices. 
The authors of  \cite{SKS}  have applied RMT to studying heat transfer by a highly connected and disordered network of oscillators. A considerable simplification occurring in \cite{SKS}, as compared to the systems discussed in the present Letter, is that the mass matrix $\M{M}$ is simply proportional to the unit matrix. Thus, these authors needed only to apply standard RMT techniques to analyze the random matrix $\M{K}$. 

At the next level of complexity lies the analysis carried in \cite{Fyodorov} of the spectral statistics of real symmetric random matrix pencils with a deterministic diagonal metric, with nice application to fully connected electrical $LC$-networks. (See also \cite{Marchenko-Pastur, Pastur} for earlier work on such pencils.)

The methods tailored for disordered chains or crystals \cite{Dyson, Schmidt, Mattis}, as well as the more standard RMT methods used in \cite{SKS},  are inapplicable for determining the average phonon (or vibrational) spectra of systems described by full non-commuting random matrices $\M{M}$ and $\M{K}$. This requires a different approach: 

Since there is no reason to expect any statistical correlation between these two matrices, we shall draw them from two independent probability ensembles. The matrix elements of either $\M{M}$ or $\M{K}$ cannot be distributed independently. The elements of each matrix are correlated by the fact that these matrices are positive. By definition, these matrices are also real. The least biased way to fulfil these constraints is to take these matrices to be of Wishart form $\M{C}^T\M{C}$, with $\M{C}$ an $N\times N$ real Ginibre matrix\cite{Ginibre}. We shall however henceforth relax the constraint that $\M{M}$ and $\M{K}$ be real, and take them to be positive hermitian, with $\M{C}$ drawn from the complex Ginibre probability ensemble
\begin{equation}\label{ginibre}
P_\sigma(\M{C})  = \frac{1}{\cal Z} e^{-{N\over\sigma^2}{\rm Tr}\,\M{C}^\dagger\M{C}}\,,
\end{equation}
with the variance tuned such that the eigenvalues of $\M{C}$ are spread uniformly in a disk of finite radius $\sigma$ in the complex plane, as $N$ tends to infinity. Here ${\cal Z}$ is a normalization constant. 
Thus, we form two such independent complex Ginibre ensembles, one for $\M{K} = \M{C}_1^\dagger\M{C}_1$ with variance $\sigma_K^2$, and another for 
$\M{M} = \M{C}_2^\dagger\M{C}_2 + m_0 $ with variance $\sigma_M^2$. The {\em positive} shift parameter $m_0$ ensures that $\M{M}\geq m_0$ is strictly positive with probability one.  

\begin{figure}[tb]
\includegraphics[width=\columnwidth]{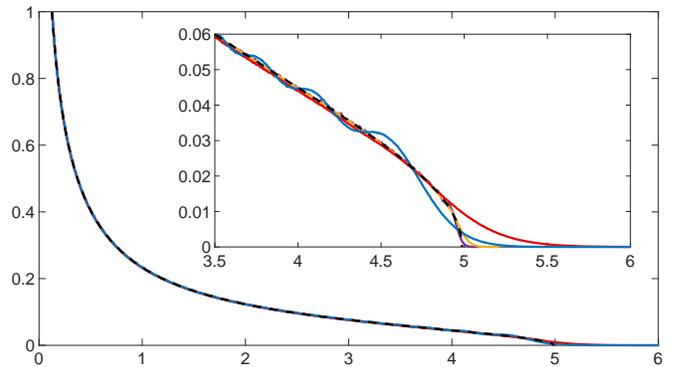}
\caption{Histogram envelope curve showing the density of eigenvalues $\varrho_{\M{H}}(\omega^2)$ of the mechanical system for complex matrices with $N=64$ (blue) and $N=1024$ (purple) and for real matrices with $N=64$ (red) and $N=1024$ (yellow), calculated from $2\times 10^7$ samples ($N=64$) and $50000$ samples ($N=1024$). The parameters of the system are $\mu=m_0=0.5$, $\sigma_M=\sigma_K=1$ (that is, $\omega_0^2=1$).  The dashed black line shows the theoretical large $N$ prediction for complex matrices given by \eqref{eq:rho}.}\label{fig:real_vs_complex}
\end{figure}

Such a generalization from real into complex matrices should not change the vibration spectrum in the thermodynamic limit. We have verified this expectation numerically: The difference between real and hermitian matrices amounts only to small finite-$N$ corrections at the high frequency band-edge, which vanish as $N$ tends to infinity (see Fig.~\ref{fig:real_vs_complex}). 
Of course, a more detailed investigation of the spectral statistics of such systems, such as studying level-spacings, will depend on whether one is studying real or complex matrices. In this Letter we focus exclusively on the average spectrum in the thermodynamic limit, which is not affected by taking $\M{M}$ and $\M{K}$ to be complex hermitian. Issues of more detailed spectral statistics of such systems is an open problem.

Having defined the probability ensembles for $\M{M}$ and $\M{K}$, we draw a pair of such matrices from their corresponding ensembles and compute the random matrix $\M{H} = \M{M}^{-1}\M{K}$, in accordance with \eqref{H}.  Our objective is to calculate the resolvent 
\begin{equation}\label{resolventH}
G_{\M{H}}(z;m_0,\sigma_M,\sigma_K) = \Big\langle {1\over N}{\rm Tr} {1\over z-\M{M}^{-1}\M{K}}\Big\rangle_{\sigma_M,\sigma_K}
\end{equation}
of $\M{H}$ averaged over $P_{\sigma_K}(\M{C}_1) $ and $P_{\sigma_M}(\M{C}_2)$, in the large-$N$ limit. We can then obtain the desired averaged density of eigenvalues $\varrho_{\M{H}}(\omega^2) =  \langle 1/N {\rm Tr}\, \delta(\omega^2-\M{H})\rangle_{\sigma_M,\sigma_K}$
of $\M{H}$ from (\ref{resolventH}) in the usual manner \cite{QFTNut}
\begin{equation}\label{phonon-density}
\varrho_{\M{H}}(\omega^2)  = \lim_{ \epsilon\rightarrow 0^+}\frac{1}{\pi}{\rm Im}G_{\M{H}}(\omega^2-i\epsilon;m_0,\sigma_M,\sigma_K)\,.
\end{equation}

An immediate consequence of \eqref{ginibre}-\eqref{resolventH} is that the resolvent \eqref{resolventH} obeys the scaling law
\begin{equation*}
\left({\sigma_K\over\sigma_M}\right)^2 G_{\M{H}}(z;m_0,\sigma_M,\sigma_K) = G_{\M{H}}(\zeta;\mu,1,1)  =:\Gamma(\zeta,\mu),
\end{equation*}
with rescaled variables
\begin{equation}\label{scaling-variables}
\zeta = \left(\frac{\sigma_M}{\sigma_K}\right)^2 z = \frac{z}{\omega_0^2}\quad {\rm and}\quad \mu = \frac{m_0}{\sigma_M^2}\,.
\end{equation}
Clearly, $\sigma_M^2$ has dimensions of mass, and $\sigma_K^2$ has dimensions of force per unit length. Thus, $\frac{\sigma_K^2}{\sigma_M^2} = \omega_0^2$ has dimensions of frequency squared, and \eqref{scaling-variables} simply instructs us to measure $m_0$ in units of $\sigma_M^2$ and the complex spectral parameter $z$ in units of $\omega_0^2$. For the density we have the relation
\begin{equation*}
\varrho_{\M{H}}(\omega^2; m_0,\sigma_M,\sigma_K)=\frac{1}{\omega_0^2}\varrho_{\M{H}}(x; \mu,1,1)=:\frac{1}{\omega_0^2}\varrho_{\M{H}}(x; \mu),
\end{equation*}
with $x=\omega^2/\omega_0^2={\rm Re}\zeta$.

Since the matrices $\M{M}$ and $\M{K}$ are positive definite and drawn from unitary invariant ensembles, we can apply {\em S-transform} techniques of free probability theory \cite{free,burda} to calculate the resolvent and density of eigenvalues of products like $\M{H}$, in the large-$N$ limit. In our case, it reduces the calculation of \eqref{resolventH} to solving the cubic equation 
\begin{eqnarray}\label{scaled-cubic}
\left(\zeta + \zeta^2\right) \Gamma(\zeta,\mu)^3 -\left[(2+\mu)\zeta + \mu\zeta^2\right]\Gamma(\zeta,\mu)^2 \nonumber\\
+ (1+\mu+2\mu\zeta)\Gamma(\zeta,\mu) - \mu=0,
\end{eqnarray}
where the relevant root is the one with asymptotic behavior $\Gamma(\zeta,\mu)\sim \frac{1}{\zeta}$ as $\zeta\rightarrow\infty$.
To find the density of eigenvalues we have to look for the imaginary part of $\Gamma(\zeta,\mu)$ when $\zeta=x$ is real (see Eq.~\eqref{phonon-density}). We introduce the discriminant of the cubic equation \eqref{scaled-cubic},
\begin{align*}
\Delta_\Gamma&(x)=x\Big( (\mu+4)\mu^3 x^3 +2\mu^2(\mu^2+2\mu-6) x^2 \nonumber \\
&+(\mu^3-4 \mu^2-20 \mu+12)\mu x -4(\mu^3+3\mu^2+3\mu+1) \Big).  
\end{align*} 
The discriminant $\Delta_\Gamma(x)$ has two real roots $0$ and $x_1$, 
\begin{align*}
x_1=&\frac{1}{3\mu^3(\mu+4)}\Big(-2\mu^2(\mu^2+2\mu-6) \nonumber\\
&\qquad+\frac{(\xi_1+\sqrt{\Delta_1})^{1/3} +(\xi_1-\sqrt{\Delta_1})^{1/3}}{2^{1/3}} \Big), 
\end{align*}
 with
\begin{align*}
\xi_1&=2\mu^7(\mu^5+24\mu^4+264\mu^3+1574\mu^2+4806\mu+5832),\\
\Delta_1&=432\mu^{14}(\mu+4)^2 (\mu^2+10\mu+27)^3>0.
\end{align*}
The interval $(0,\omega_0^2 x_1)$ is therefore the desired support of $\varrho_{\M{H}}(\omega^2)$. 
This support is purely positive as it should be, due to positivity of the matrix $\M{H}$.  $x_1(\mu)$ is a monotonically decreasing function, which should be expected physically because $\omega^2\sim \M{M}^{-1}\M{K}\sim 1/\mu$. 
Picking that root of the cubic \eqref{scaled-cubic} which has positive imaginary part along $[0,x_1]$ we thus find 
\begin{equation}
\varrho_{\M{H}}(x;\mu)=\tfrac{1}{2\sqrt{3} x(x+1)\pi}\left(\frac{(\xi_\Gamma+\delta_\Gamma)^{1/3}}{2^{1/3}}-\frac{2^{1/3}\chi_\Gamma}{(\xi_\Gamma+\delta_\Gamma)^{1/3}}\right),\label{eq:rho}
\end{equation}
where
\begin{align*}
\xi_\Gamma&=
-x^2 [2\mu^3 x^4 +6\mu^2(\mu-1) x^3 +3\mu(2\mu^2-7\mu+2) x^2 \nonumber \\  
&\qquad\qquad\quad+2(\mu^3-12\mu^2+3\mu-1)x -9(\mu^2+2)], \\
\delta_\Gamma&=
x(x+1)\sqrt{-27\Delta_\Gamma(x)}, \\ 
\chi_\Gamma&=
\mu^2 x^4+2\mu(\mu-1)x^3 +(\mu^2\!-5\mu+1)x^2-3(\mu+1)x. 
\end{align*}
$\varrho_{\M{H}}(x;\mu)$ diverges at the origin like $1/\sqrt{x}$, and vanishes at $x_1$ like $\sqrt{x_1-x}$. 
This divergence of $\varrho_{\M{H}}(x;\mu)$ like $1/\sqrt{x}\sim 1/\omega$ at the origin is reminiscent the behavior \eqref{universal} of pendula, indicating a {\em universal} such behavior of the density of vibration eigenmodes of highly connected systems at low frequencies. 

\begin{figure*}[tbh!]
(a)\includegraphics[width=0.98\columnwidth]{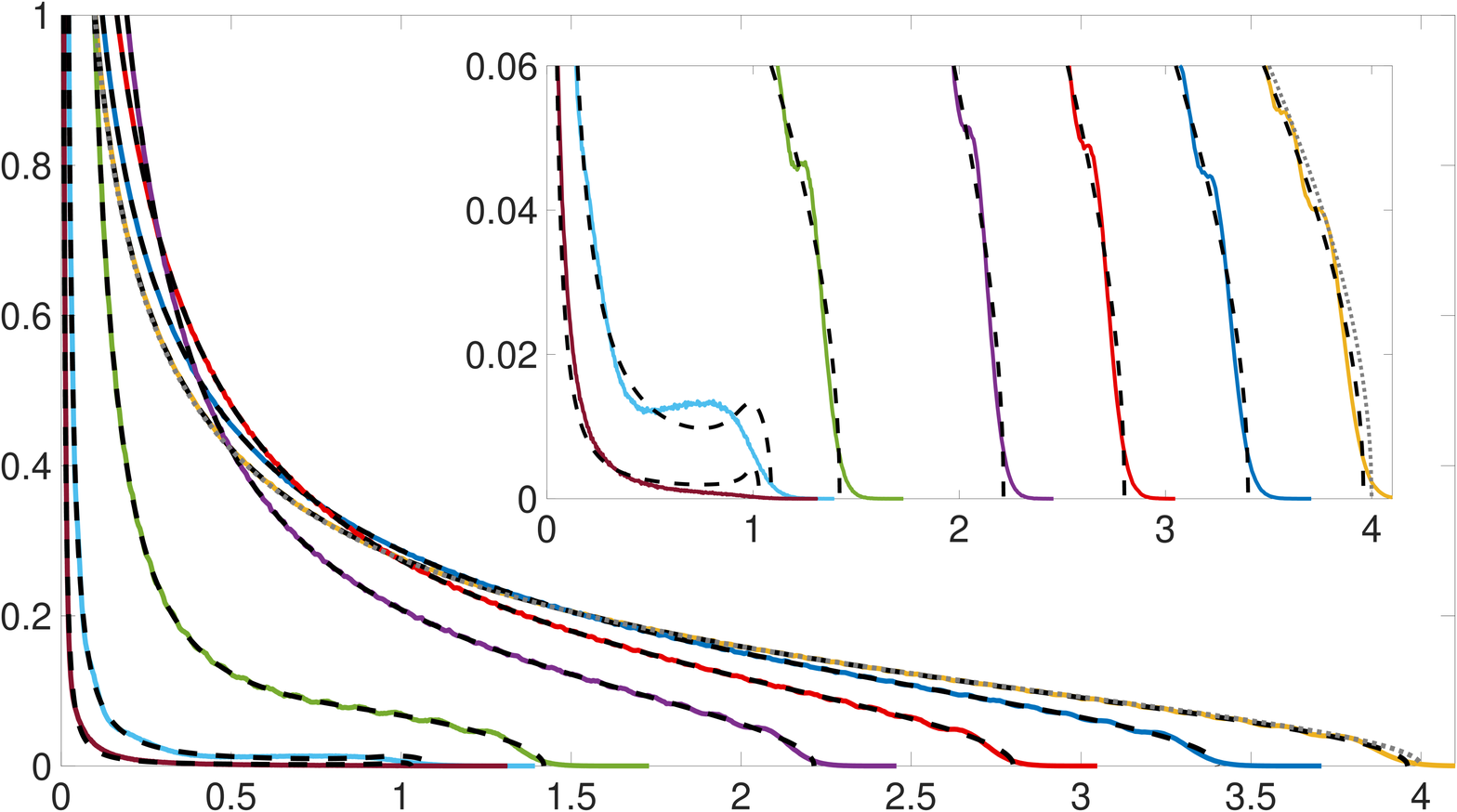}
\hfill(b)\includegraphics[width=0.98\columnwidth]{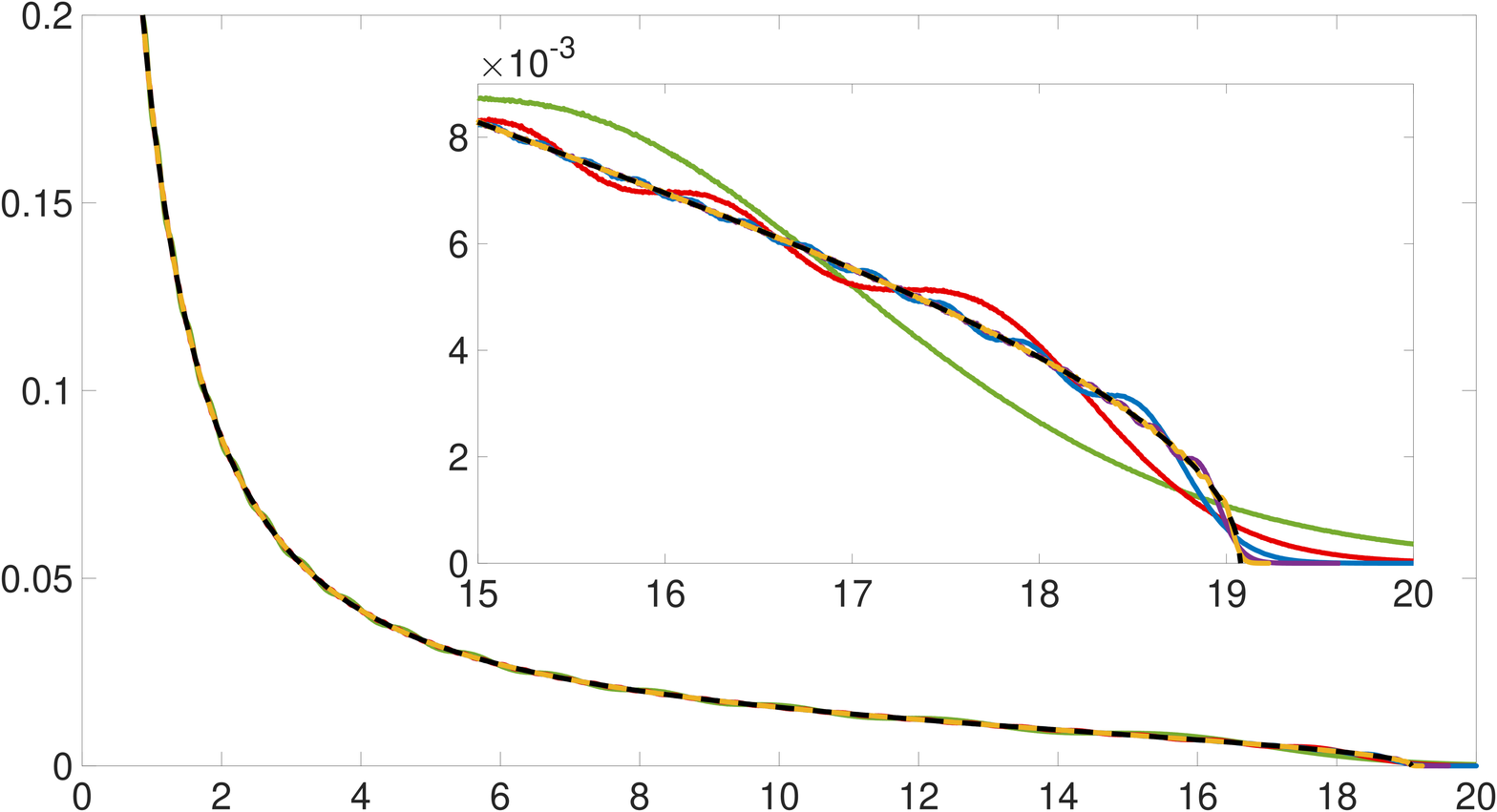}
\caption{Plot of the density of eigenvalues $\varrho_{\M{H},N}^\text{num}(\omega^2)$ from a simulation using millions of samples of complex matrices. In (a) we show it for the fixed parameters $N=128$, $\sigma_K=1$, and $m_0=1$, for various values of $\sigma_M$, in order of increasing endpoints $\omega_0^2 x_1$: 500 (brown), 100 (cyan), 10 (green), 2 (purple), 1 (red), 0.5 (blue) and 0.1 (yellow). The dotted line (close to the yellow line) shows the Marchenko-Pastur distribution. In (b) we show it for the fixed parameters $\mu=0.1$, $\sigma_M=\sigma_K=1$ ($\omega_0^2=1$), for various values of $N$: $N=32$ (green), $N=128$ (red), $N=512$ (blue), $N=2048$ (purple) and $N=8192$ (yellow). The dashed black line shows the theoretical large-$N$ prediction given by \eqref{eq:rho}.}\label{Fig:sigmaN}
\end{figure*}

Let $\varrho_{\M{H},N}^\text{num}(\omega^2)$ denote the finite $N$ averaged density of the mechanical system from numerical simulations. It is described by the curves in Fig.~\ref{Fig:sigmaN}. They are the envelope curves gleaned from histograms with very narrow bins. These curves for $\varrho_{\M{H},N}^\text{num}(\omega^2)$ are in excellent agreement with the analytical expression for $\varrho_{\M{H}}(\omega^2)$ in \eqref{eq:rho} when $N$ is large. In Fig.~\ref{Fig:sigmaN}a we display $\varrho_{\M{H},N}^\text{num}$ obtained by averaging over a million samples, and compare it to $\varrho_{\M{H}}$, for various values of the parameter $\sigma_M$. Note that in the limit $1/\mu\rightarrow 0$, $\varrho_{\M{H},N}^\text{num}$ converges to the Marchenko-Pastur density as should be expected, because in this limit the matrix $\M{M}$ becomes deterministic and proportional to the unit matrix. In the opposite limit of large $\sigma_M$ (or equivalently $\mu\rightarrow 0$), the density looks qualitatively different from the Marchenko-Pastur profile.  

In Fig.~\ref{Fig:sigmaN}b we show numerical results for the density, with a fixed choice of parameters and for various values of $N$. Convergence of the numerical results to the theoretical large-$N$ curve in the bulk is rapid, whereas convergence close to the high frequency (soft) edge is non-uniform, with visible finite-$N$ corrections. Since we have used complex matrices, the latter corrections clearly exhibit oscillations towards the high frequency edge and we have verified that they match with the Airy behavior of the canonical GUE case. On the other hand, the model with real matrices has non-oscillatory edge behavior, like in the GOE case (as can be seen in Fig.~\ref{fig:real_vs_complex}).

{\it Eigenvectors and their Participation Ratio.} Consider the \emph{normalized} eigenvector $\M{A}(\omega^2)$ of $\M{H}$ in \eqref{eigen}, corresponding to eigenvalue $\omega^2$ with components $A_\ell$. 
The (normalized) \emph{participation ratio}\cite{wegner} is the function of $\omega^2$
\begin{equation}\label{eq:PR}
p(\omega^2)=\frac{1}{N}\Big\langle\frac{1}{\sum_{\ell=1}^N |A_\ell(\omega^2)|^4}\Big\rangle_{\sigma_M,\sigma_K}. 
\end{equation}
It is a measure of the fraction (out of $N$) of degrees of freedom of the system that are effectively involved in a given state of vibration. We have computed numerically the participation ratios for the complex and real matrix models and found that these participation ratios are independent of the eigenvalue $\omega^2$ and converge to constants as $N$ increases. This means that in both models all states are extended. That is, in all vibrational modes, essentially all degrees of freedom oscillate with amplitudes of the same order of magnitude. In other words, vibration eigenmodes tend to be collective throughout the frequency band. Moreover, these constant values seem to be universal (for various values of $\mu$) and approach (for large $N$) the value of 0.50 in the complex case and 0.33 for the real model. Even for low $N$ these numerical results coincide, respectively, with the participation ratios of the canonical GUE and GOE (obtained from the Porter-Thomas probability distributions \cite{haake}).

\begin{figure}[b]
	\includegraphics[width=\columnwidth]{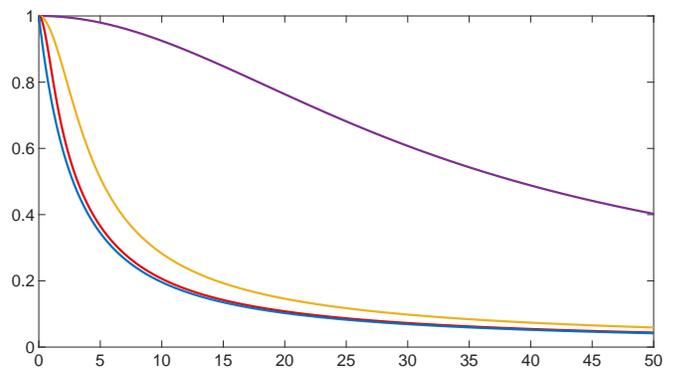}
	\caption{Plot of the  the specific heat $c_V/k_\text{B}$ given by \eqref{intensive} against inverse temperature $\beta$ (for $\hbar=1$). The parameters of the model are $\sigma_M=\sigma_K=1\, (\omega_0^2 = 1)$ and $\mu=10^{-4}$ (blue), $\mu=0.1$ (red) $\mu=1$ (yellow) and $\mu=100$ (purple).}\label{Fig:CV}
\end{figure}

{\it Statistical Mechanics of Phonons.} 
Upon quantization, the normal modes of our system amount to a collection of non-interacting quantum harmonic oscillators. In a state of thermal equilibrium at temperature $T$, the average energy tied with the oscillator with frequency $\omega$ is  
\begin{equation}\label{mode-energy}
\bar {\cal E}(\omega,T) = \hbar\omega\left(\frac{1}{2} + {1\over e^{\beta\hbar\omega}-1}\right)\,, 
\end{equation}
where $\beta = \frac{1}{k_\text{B} T}$. 
Thus, the ensemble average total thermal energy is 
\begin{equation}\label{energy1}
\Big\langle \bar E(T)\Big\rangle_{\sigma_M,\sigma_K} = N\int\limits_0^\infty \bar{\cal E}(\omega,T)2\omega\varrho_{\M{H}}(\omega^2)d\omega\,.
\end{equation}
This is, of course, an extensive quantity, proportional to $N$. 
Therefore, the specific heat (per degree of freedom) is,  
\begin{eqnarray}\label{intensive}
c_V(T) &=& \frac{k_\text{B}(\beta \hbar)^2}{2} \! \int_0^\infty \!\! \frac{\omega^3 \varrho_{\M{H}}(\omega^2)}{\sinh^2(\beta\hbar\omega/2)} d\omega.
\end{eqnarray}
Results from numerical integration of \eqref{intensive} over the explicit expression for $\varrho_{\M{H}}(\omega^2)$ in Eq.~\eqref{eq:rho} are displayed in Fig.~\ref{Fig:CV}.
At large temperatures $c_V/k_\text{B}$ goes to 1 (the classical limit) while the energy \eqref{energy1} goes to $N k_\text{B} T$ (classical equipartition). High temperature means that $k_\text{B}T\gg \hbar \omega_\text{max}$ where $\omega_\text{max}=\sqrt{x_1}$. As $T\rightarrow 0$ all oscillation modes become frozen at their ground states with zero-point energy (ZPE) $\hbar \omega/2$. Thus the zero temperature limit of the energy is just the spectral sum over all ZPE up to exponentially small corrections $\sim e^{-\beta\hbar\omega}$. Consequentially $c_V$ is exponentially small and gets most of its contribution from the low frequency part of the spectrum. 
For large $\mu$ (see the purple plot in Fig.~\ref{Fig:CV}), as we discussed earlier, the spectral density tends to the Marchenko-Pastur profile, which has significant spectral weight at small frequencies. Thus, $c_V$ decays very slowly as a function of $\beta$ for large values of $\mu$.

{\it Universal behavior for small frequencies.}
An important result of this Letter is that the density of eigenfrequencies $\tilde\rho(\omega)= 2\omega \varrho_{\M{H}}(\omega^2)$ of both our matrix model and pendula tend to a nonvanishing constant in the limit of small frequencies, which seems to be a common universal feature of highly connected systems. Low frequency modes are long-wavelength collective vibration modes, and they are expected to probe the mechanical system as a whole, in some sense. Thus, it is quite surprising that the density of these modes in our highly connected systems is qualitatively the same as that of acoustic phonons in one-dimensional perfect crystals, with its nearest-neighbor interatomic interactions, which is also constant. 
More formally, if we think of our ''hamiltonian`` $\M{H}$ as the discrete laplacian of some graph associated with our highly connected mechanical system, then the {\it spectral dimension} $d_S$ of that graph is defined by the scaling behavior $\varrho_{\M{H}}(\omega^2)\sim (\omega^2)^{d_S/2-1}$ for $(\omega/\omega_0)^2 \ll 1$ \cite{HbA}, familiar from the theory of diffusion on fractal graphs \cite{AB,RT} (see also \cite{ADT}). Thus for our system, indeed $d_S=1$, which seems to be a {\it universal} feature of vibrational spectra of highly connected systems. Spectral dimension $d_S=2$ seems to correspond to the vibrational spectrum of globular proteins \cite{bA}. For a very recent discussion of spectral dimensions in the context of complex networks see \cite{Bianconi}.

\begin{acknowledgments}
This research was supported by the Israel Science Foundation (ISF) under grant No. 2040/17.  Computations presented in this work were performed on the Hive computer cluster at the University of Haifa, which is partly funded by ISF grant 2155/15. JF wishes to thank Andreas Fring, Tsampikos Kottos and Boris Shapiro for valuable discussions and for suggesting several references. 
\end{acknowledgments}

\end{document}